\documentclass[10pt,a4paper]{article}

\usepackage[T1]{fontenc}
\usepackage[utf8]{inputenc}
\usepackage{textcomp}
\usepackage{libertine}
\usepackage[libertine]{newtxmath}
\usepackage[varqu,varl,scaled=0.86]{zi4}   

\usepackage[a4paper,top=2.0cm,bottom=2.0cm,left=2.1cm,right=2.1cm,
            headheight=14pt,footskip=1.0cm]{geometry}

\usepackage{array}
\usepackage{booktabs}
\usepackage{longtable}
\usepackage{tabularx}
\usepackage{makecell}
\usepackage{amsmath}
\usepackage{xcolor}

\usepackage{graphicx}
\usepackage{tikz}
\usepackage{pgfplots}
\pgfplotsset{compat=1.18}
\usepgfplotslibrary{groupplots}
\usetikzlibrary{arrows.meta,positioning,shapes.geometric,shapes.misc,calc,
                fit,backgrounds,decorations.pathreplacing,patterns}

\usepackage{tcolorbox}
\tcbuselibrary{breakable,skins}

\usepackage{listings}
\usepackage{xspace}

\usepackage{microtype}
\usepackage{enumitem}
\usepackage{fancyhdr}
\usepackage{titlesec}
\usepackage{placeins}
\usepackage{xurl}
\usepackage{cite}

\definecolor{cH100}{RGB}{150,166,196}   
\definecolor{cH200}{RGB}{104,132,178}   
\definecolor{cB200}{RGB}{214,158,124}   
\definecolor{cB300}{RGB}{196,112,74}    
\definecolor{cNeg}{RGB}{160,60,60}      
\definecolor{cKey}{RGB}{70,100,140}     
\definecolor{cFabric}{RGB}{196,112,74}  
\definecolor{cChip}{RGB}{104,132,178}   
\definecolor{cHost}{RGB}{120,120,120}   
\definecolor{linkblue}{RGB}{20,70,130}
\definecolor{codebg}{RGB}{247,247,244}

\usepackage[hidelinks,bookmarksnumbered=true]{hyperref}

\usepackage[font=footnotesize,labelfont=bf,skip=3pt]{caption}
\pgfplotsset{
  figbase/.style={
    width=\linewidth, height=3.9cm,
    tick label style={font=\scriptsize},
    label style={font=\scriptsize},
    title style={font=\scriptsize\bfseries},
    legend style={font=\scriptsize, draw=none, fill=none, inner sep=1pt},
    every axis plot/.append style={line width=0.8pt},
  },
  figwide/.style={
    figbase, width=0.47\textwidth, height=3.8cm,
  },
}

\newcommand{\fitwidth}[1]{%
  \sbox0{#1}%
  \ifdim\wd0>\dimexpr\linewidth-1pt\relax
    \makebox[\linewidth][l]{\resizebox{\linewidth}{!}{\usebox0}}%
  \else
    \usebox0
  \fi}

\newcommand{\sys}{\textsf{flash-comm}\xspace}

\newcommand{\code}[1]{\texttt{\small #1}}

\titleformat{\section}
  {\normalfont\Large\bfseries}{\thesection}{0.7em}{}
\titleformat{\subsection}
  {\normalfont\large\bfseries}{\thesubsection}{0.6em}{}
\titlespacing*{\section}{0pt}{2.6ex plus 1ex minus .2ex}{1.5ex plus .2ex}
\titlespacing*{\subsection}{0pt}{2.6ex plus 1ex minus .2ex}{1.0ex plus .2ex}
\titleformat{\subsubsection}
  {\normalfont\normalsize\bfseries}{\thesubsubsection}{0.6em}{}
\titlespacing*{\subsubsection}{0pt}{1.8ex plus .8ex minus .2ex}{0.6ex plus .2ex}

\newcommand{\appendixformat}{%
  \titleformat{\section}
    {\normalfont\Large\bfseries}{Appendix \thesection}{0.7em}{}}

\newtcolorbox{kp}[1][]{%
  breakable, enhanced jigsaw,
  colback=cKey!6, colframe=cKey!55,
  boxrule=0pt, leftrule=2.6pt, arc=1pt, outer arc=1pt,
  left=8pt, right=8pt, top=6pt, bottom=6pt,
  fontupper=\normalsize, #1}

\newtcolorbox{warnbox}[1][]{%
  breakable, enhanced jigsaw,
  colback=cNeg!5, colframe=cNeg!50,
  boxrule=0pt, leftrule=2.6pt, arc=1pt, outer arc=1pt,
  left=8pt, right=8pt, top=6pt, bottom=6pt,
  fontupper=\normalsize, #1}

\newtcolorbox{takeaway}[1][]{%
  breakable, enhanced jigsaw,
  colback=black!3, colframe=black!45,
  boxrule=0pt, leftrule=2.6pt, arc=1pt, outer arc=1pt,
  left=8pt, right=8pt, top=6pt, bottom=6pt,
  fontupper=\normalsize, #1}

\lstdefinestyle{cuda}{
  language=C,
  basicstyle=\ttfamily\footnotesize,
  commentstyle=\color{cKey!85!black}\itshape,
  keywordstyle=\color{cB300!80!black}\bfseries,
  stringstyle=\color{cNeg!80!black},
  backgroundcolor=\color{codebg},
  frame=none, framesep=6pt,
  xleftmargin=0pt, xrightmargin=0pt,
  aboveskip=1.0em, belowskip=1.0em,
  showstringspaces=false,
  breaklines=true,
  columns=fullflexible,
  keepspaces=true,
  upquote=true,
  morekeywords={__device__,__forceinline__,uint4,uint32_t,constexpr,
                template,class,typename,asm,volatile,uint64_t},
}
\begin{document}

\begin{center}
{\LARGE\bfseries flash-comm: Self-Synchronizing Collective Communication\par}
\vspace{0.8em}
{\Large\bfseries Efficient Transport for Distributed LLM Decoding\par}
\vspace{1em}
\begin{tabular}{@{}c@{\hspace{2.5em}}c@{\hspace{2.5em}}c@{}}
Carrie Wang & Jack Li & Jason Zhao \\
\href{http://ScitiX.AI}{ScitiX.AI} &
\href{http://ScitiX.AI}{ScitiX.AI} &
\href{http://ScitiX.AI}{ScitiX.AI} \\
\end{tabular}
\end{center}



\section*{Abstract}

Collective communication, which synchronizes data across a group of GPUs, 
    lies on the critical path of decoding in distributed large language model (LLM) serving.
Despite advances in GPU interconnects and communication libraries, 
    these operations still incur substantial overhead. 
Our measurements reveal that inter-GPU barriers,
    which are used to coordinate data readiness,
    can cost more than the data transfer itself,
    yet this overhead is often overlooked in existing communication optimizations.

We present \sys, 
    a communication library that achieves efficient transport for LLM decoding 
    by \emph{fusing readiness synchronization with data movement}. 
\sys introduces \emph{self-synchronizing transfers} 
    in which each data block carries evidence of its readiness, 
    allowing receivers to directly validate the blocks they need 
    without separate inter-GPU barriers.
Based on this principle, 
    a cost-guided protocol selector adapts the readiness encoding 
    to message size and the number of participating GPUs,
    balancing synchronization overhead against additional data-transfer cost.
\sys further \emph{decouples transport order from reduction order} to
    preserve result consistency across GPUs under asynchronous data arrivals.

On a server with eight NVIDIA B300 GPUs, 
    \sys reduces all-reduce latency by a median of 28.8\% and up to 43.7\% 
    compared with custom AR v2 (the state-of-the-art all-reduce implementation). 
In end-to-end decoding, \sys reduces time per output token (TPOT) by 1.7--6.8\%.
Note that all-reduce accounts for 11.4\% of the baseline decoding time, 
    which bounds the latency reduction possible from optimizing this operation alone.
The same mechanism also supports standalone all-gather, 
    reducing latency by a median of 23.8\% and up to 58.9\% across eight B300 GPUs.


\vspace{0.8em}
\noindent\rule{\linewidth}{0.4pt}
\vspace{0.4em}

\section{Introduction}
\label{sec:intro}

Distributed large language model (LLM) serving partitions computation across
GPUs, which must exchange intermediate results before dependent computation
can proceed. During decoding, these exchanges take the form of collective
operations that recur across layers and generated tokens, so even
microsecond-scale overhead accumulates on the critical path. Faster
interconnects and optimized communication libraries have reduced this cost,
yet it remains substantial: in a profiled workload, all-reduce alone accounts
for 11.4\% of the baseline decoding time. Further reducing communication
overhead is therefore important for lowering time per output token (TPOT).

To understand where this overhead comes from, we examine the synchronization
needed to exchange data safely. Inter-GPU barriers establish when data is
ready to read and when shared buffers can be reused, but incur fixed
coordination costs even for short transfers. On B300, for example, a
cross-GPU barrier costs
4.33\,$\mu$s, compared with 3.33\,$\mu$s for an entire 1\,KB
all-reduce. To eliminate this barrier overhead, we present \sys, a
communication library for LLM decoding that \emph{fuses readiness
synchronization with data movement}. Its \emph{self-synchronizing transfers}
make each data block carry evidence of its readiness, so receivers can
directly validate the blocks they need. This allows communication to progress
without separate inter-GPU barriers while preserving data dependencies and
safe buffer reuse.

\begin{figure}[t]
\centering
\includegraphics[width=\linewidth]{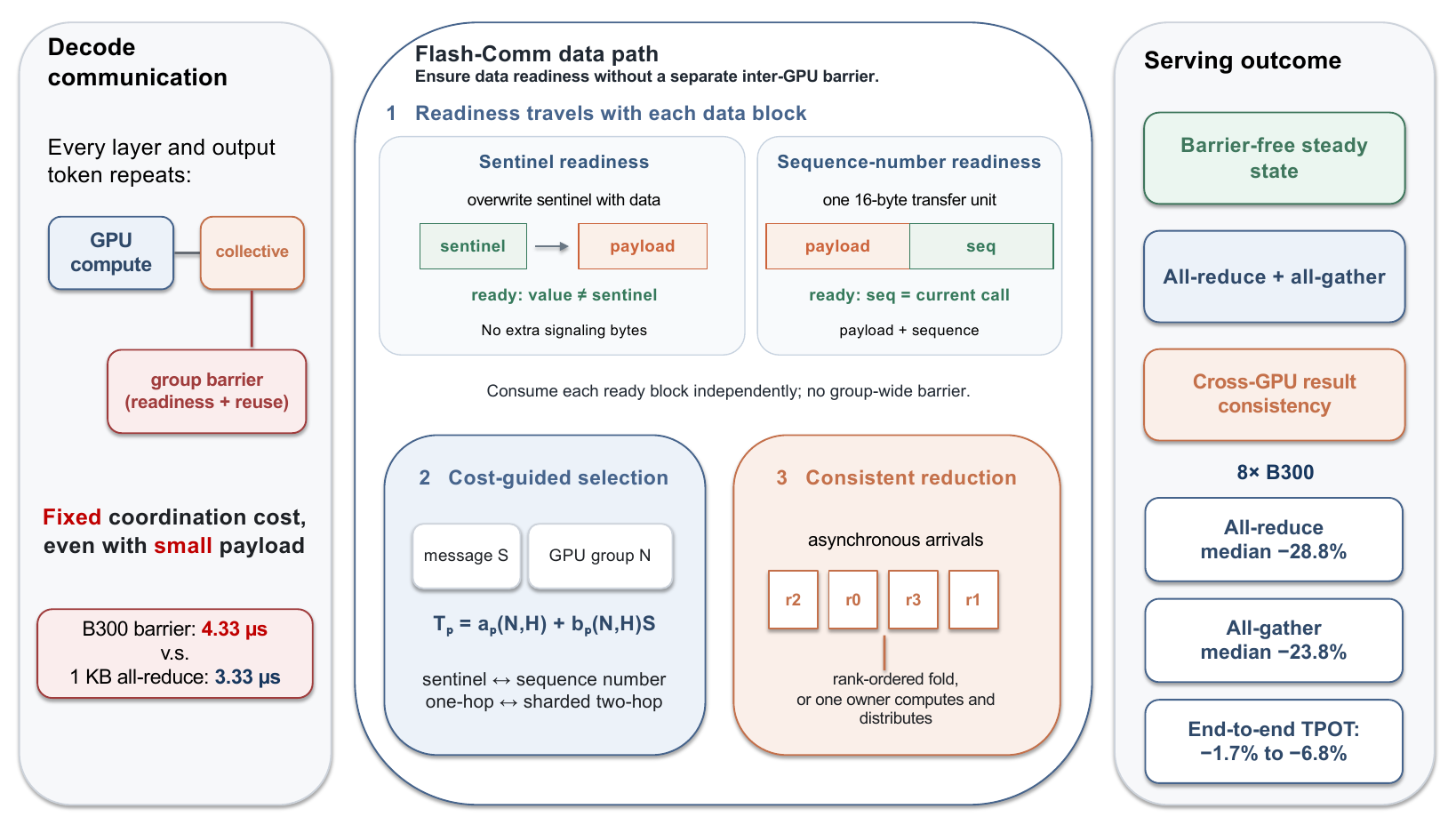}
\caption{Overview of \sys.}
\label{fig:system-overview}
\end{figure}

Figure~\ref{fig:system-overview} summarizes how these ideas fit together.
\sys replaces group-wide readiness coordination with block-level readiness
evidence, selects the readiness encoding and communication path according to
their fixed and per-byte costs, and applies a deterministic reduction order
when asynchronously delivered values are combined.

Realizing this design efficiently requires choosing how to encode readiness.
Encoding it in data values avoids extra transmitted bytes but requires
checking and resetting the buffers, whereas explicit sequence numbers
simplify these operations at the cost of additional traffic. The balance
depends on the communication workload: extra bytes matter less for short
messages, but become more costly as transfer volume grows. To account for
this tradeoff, \sys uses a \emph{cost-guided protocol selector} that separates
fixed overhead from per-byte cost. Measurements calibrate the switching
boundaries for different message sizes and GPU configurations.

Beyond this performance tradeoff, allowing transfers to progress asynchronously
raises a correctness concern for reductions. Because floating-point addition
is not associative, GPUs that accumulate contributions in arrival order can
produce different results from the same inputs. To preserve consistency,
\sys \emph{decouples transport order from reduction order}. When GPUs compute
the same output independently, they accept contributions in any arrival order
but accumulate them in a common rank order. For sharded reductions, a single
GPU computes each output block and distributes the same result to the others.
In both cases, consistency does not require serializing data arrival.

We evaluate these mechanisms through collective communication benchmarks and
end-to-end decoding workloads. On a server with eight NVIDIA B300 GPUs,
\sys reduces all-reduce latency by a median of 28.8\% and up to 43.7\%
compared with custom AR v2. Applying the same transfer mechanism to standalone
all-gather reduces latency by a median of 23.8\% and up to 58.9\% across
eight B300 GPUs. These communication improvements also translate to lower
token generation latency: in end-to-end decoding, \sys reduces TPOT by
1.7--6.8\%.

\section{Background and Motivation}
\label{sec:background}

\subsection{Collective Communication in LLM Decoding}
\label{sec:decode-communication}

Distributed LLM serving partitions model execution across multiple GPUs.
Each GPU computes only part of an operator, so the devices must synchronize
their intermediate data before dependent computation can proceed. These data
exchanges are expressed as collective operations. All-reduce combines partial
results and returns the result to every GPU; all-gather assembles distributed
pieces on every GPU; and all-to-all sends each piece to its designated
destination. Together, these operations support common serving strategies
such as tensor parallelism and expert parallelism.

Communication is particularly visible during autoregressive decoding. Each
decoding step produces only the next token or a short speculative sequence,
but repeats the same communication dependencies across layers and generated
tokens. A few microseconds added to one collective are therefore paid many
times along the token-generation path. The message size varies with the model,
batch size, data type, and parallel execution strategy, spanning transfers
where fixed protocol overhead dominates and transfers where data movement
dominates. In a representative profiled workload, all-reduce takes 1.68\,ms of
a 14.73\,ms decoding iteration, or 11.4\% of the baseline time.
Collective communication therefore remains a meaningful optimization target
even in an optimized serving stack.

\subsection{Data Readiness and Buffer Reuse}
\label{sec:readiness-background}

Fast links do not by themselves tell a receiver when remote data is safe to
consume. A GPU may enter its communication kernel while a peer is still
producing its input, or before that peer's writes are visible across the
interconnect. Reading the shared buffer at that point can return data from the
previous invocation or an incompletely published value. A communication
protocol must therefore establish that the producer has finished writing the
required data and that the consumer can observe those writes.

The reverse dependency governs buffer reuse. A producer cannot overwrite a
shared region for the next invocation while another GPU may still be reading
the current value. Multi-stage collectives add a dependency between phases.
For example, an all-reduce implemented as reduce-scatter followed by
all-gather cannot distribute a reduced block until that block is complete.
Correct execution thus requires three related guarantees: data is published
before it is consumed, one phase completes before its dependent phase starts,
and storage is reused only after the relevant consumers are done.

Inter-GPU barriers are a common way to establish these relationships for the
whole group. Together with the required memory-ordering operations, a barrier
places every participant on the same side of a protocol boundary before
dependent reads or buffer reuse. This group-wide confirmation is sufficient,
but it is stronger than the immediate dependency of a consumer, which needs
to know only whether its particular data block is ready. If each block carries
reliable evidence of readiness, a receiver can validate the blocks it needs
directly. Buffer reuse can likewise follow evidence that the relevant value
has been consumed. This observation motivates the self-synchronizing transfers
in \sys: preserve the data dependencies, but establish them through the data
path instead of a separate group-wide confirmation step.

\subsection{The Cost of Separate Synchronization}
\label{sec:separate-synchronization}

A barrier carries little payload, yet it still requires participants to
publish and observe state across GPUs. Its latency is therefore largely a
fixed protocol cost and does not shrink with the message being exchanged. We
measure this cost under CUDA Graph execution using a flat cross-GPU barrier,
then subtract the empty-kernel floor to isolate the inter-GPU component.
Table~\ref{tab:barrier-cost} compares that net cost with the latency of a
complete 1\,KB all-reduce on the same machine. 
As shown in the table, the standalone barrier costs more than the entire small-message
all-reduce on B300.

\begin{table}[t]
\centering
\caption{Net cross-GPU barrier latency is comparable to the latency of a
complete 1\,KB all-reduce. Both are measured under CUDA Graph execution.}
\label{tab:barrier-cost}
\begin{tabular}{@{}l r r@{}}
\toprule
\textbf{Machine} & \textbf{Net barrier latency} &
  \textbf{1\,KB all-reduce latency} \\
\midrule
H100 & 2.66\,$\mu$s & 2.54\,$\mu$s \\
H200 & 1.87\,$\mu$s & 2.46\,$\mu$s \\
B300 & 4.33\,$\mu$s & 3.33\,$\mu$s \\
\bottomrule
\vspace{5pt}
\end{tabular}
\end{table}


These measurements show that establishing readiness can cost as much as the
small collective itself. A receiver fundamentally needs to wait only for the
data on which it depends, whereas a barrier adds a group-wide coordination
round to this local dependency. Block-level readiness instead lets the
receiver proceed as soon as its required data arrives. \sys follows this
approach and couples safe consumption and buffer reuse to the same readiness
evidence. The next sections present the encodings and data paths that realize
this principle efficiently.

\section{Flash-Comm Design}
\label{sec:channels}
\label{sec:design}

\sys is built around one principle: readiness synchronization should travel
with the data whose readiness it establishes. This turns synchronization from
a group-wide protocol phase into a property that a receiver checks for each
data block. Two further decisions make the principle practical. First, the
readiness encoding and communication path are selected according to their
fixed and per-byte costs. Second, reduction follows a deterministic evaluation
order even when transport is asynchronous. This section presents these three
parts of the design.

\subsection{Self-Synchronizing Transfers}
\label{sec:self-sync}

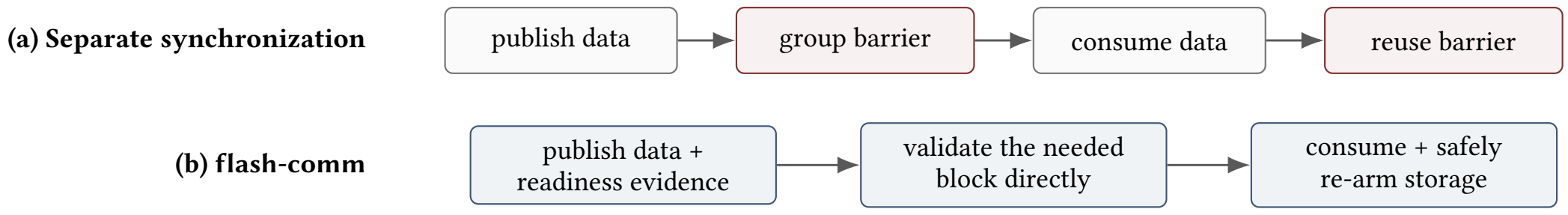
\begin{figure}[t]
\centering
\begin{tikzpicture}[
  x=1cm,y=1cm,
  stagebox/.style={draw=black!55, rounded corners=2pt, fill=black!2,
               minimum height=0.58cm, align=center, font=\scriptsize,
               inner xsep=7pt},
  sync/.style={stagebox, draw=cNeg!80!black, fill=cNeg!7},
  fused/.style={stagebox, draw=cKey!85!black, fill=cKey!8},
  flow/.style={-{Latex[length=2mm]}, line width=0.55pt, draw=black!65},
  lab/.style={font=\scriptsize\bfseries, anchor=east}
]
\node[lab] at (-0.15,1.05) {(a) Separate synchronization};
\node[stagebox, minimum width=2.05cm] (pub1) at (1.45,1.05) {publish data};
\node[sync, minimum width=2.10cm] (bar1) at (4.05,1.05) {group barrier};
\node[stagebox, minimum width=2.05cm] (use1) at (6.65,1.05) {consume data};
\node[sync, minimum width=2.10cm] (bar2) at (9.25,1.05) {reuse barrier};
\draw[flow] (pub1) -- (bar1);
\draw[flow] (bar1) -- (use1);
\draw[flow] (use1) -- (bar2);

\node[lab] at (-0.15,-0.05) {(b) \sys};
\node[fused, minimum width=2.70cm] (pub2) at (2.00,-0.05)
  {publish data +\\readiness evidence};
\node[fused, minimum width=2.70cm] (check2) at (5.45,-0.05)
  {validate the needed\\block directly};
\node[fused, minimum width=2.70cm] (use2) at (8.90,-0.05)
  {consume + safely\\re-arm storage};
\draw[flow] (pub2) -- (check2);
\draw[flow] (check2) -- (use2);
\end{tikzpicture}
\caption{\textbf{Readiness travels with each data block.}
Unlike a barrier-based protocol, \sys lets each receiver validate and consume
a block as soon as it is ready, then safely reuse its storage.}
\label{fig:self-sync}
\end{figure}

Figure~\ref{fig:self-sync} contrasts a barrier-based transfer with \sys. In a
barrier-based protocol, producers publish data and then enter a separate
group-wide synchronization step. Consumers wait for that step before reading,
even when the particular block they need is already available. A second
coordination point may be needed before buffers can be reused. In \sys, the
published block carries its own readiness evidence. A consumer polls the
block, consumes it as soon as it is ready, and re-arms or advances the storage
according to that block's consumption rule. Readiness evidence directly orders
consumption and storage reuse at block granularity. \sys provides two
mechanisms for carrying this evidence: sentinel readiness and sequence-number
readiness.

\textbf{Sentinel readiness.} The primary encoding reserves a bit pattern as a
sentinel and pre-fills receive storage with it. A producer publishes a data
pack by overwriting the sentinel. Observing a non-sentinel pack therefore
proves that the pack has arrived, without adding a flag to the transmitted
payload. After the sole consumer has captured the value, it restores the
sentinel where the buffer-reuse rule permits.

All fp32 and bf16 bit patterns are valid inputs, so a real value can collide
with the sentinel. \sys prevents such a collision from looking permanently
unready by replacing that exact pattern with a canonical quiet NaN before
publication. The same sanitization is applied to partial sums because a
reduction can create the sentinel pattern even when no individual input has
it. Legitimate NaN and infinity values create a second ambiguity in a
switch-side reduction: a non-finite result may mean either ``an input is still
the sentinel'' or ``the true result is non-finite.'' After bounded polling,
\sys verifies each source explicitly and completes the reduction once every
source has published.

\textbf{Sequence-number readiness.} An alternative encoding places data and a
call sequence number in the same 16-byte transfer unit. A receiver accepts the
data only when the sequence number matches the current invocation. This avoids
sentinel tests and buffer re-arming, but only half of each transfer unit holds
payload, doubling the traffic for the same useful data.

Both encodings operate over a set of rotating buffer slots. The slot is chosen
by a device-side call counter so that CUDA Graph replay does not depend on a
host parameter frozen during capture. Slot rotation delays reuse, while the
readiness protocol establishes whether the value in a slot belongs to the
current operation. Together, they separate successive invocations while
keeping every rank aligned on the same collective sequence.

\subsection{Cost-Guided Protocol Selection}
\label{sec:protocol-selection}
\label{sec:standing}

No readiness encoding or communication path is best for every workload.
Sentinels carry no additional payload but require tests and re-arming stores;
sequence numbers reduce that fixed work but increase the number of bytes.
Likewise, a one-hop all-reduce has one communication phase but sends a full
contribution toward every peer, whereas a sharded two-hop path adds a phase
and reduces the data volume handled by each GPU.

Within a stable protocol band, \sys models latency as
\begin{equation}
  T_p(S;N,H) = a_p(N,H) + b_p(N,H)S,
  \label{eq:cost-model}
\end{equation}
where $S$ is message size, $N$ is the number of participating GPUs, and $H$
identifies the hardware. The intercept $a_p$ includes launch, readiness
handling, and protocol round trips; the slope $b_p$ captures per-byte transfer
and memory cost. We calibrate the fixed and per-byte costs for each target
platform. The model exposes the central tradeoff: reducing fixed work
may add bytes, while reducing bytes may add a phase.
Table~\ref{tab:protocol-tradeoffs} summarizes this tradeoff for the two
readiness encodings.

\begin{table}[t]
\centering
\caption{\textbf{The two readiness encodings trade fixed protocol work for
transferred bytes.} The selector chooses between them according to the
communication workload and measured hardware costs.}
\label{tab:protocol-tradeoffs}
\fitwidth{%
\begin{tabular}{@{}p{0.24\linewidth} p{0.28\linewidth} p{0.25\linewidth}
                p{0.17\linewidth}@{}}
\toprule
\textbf{Choice} & \textbf{Reduces} & \textbf{Adds} &
\textbf{Favoured when} \\
\midrule
Sentinel readiness & transmitted signaling bytes & checking and re-arming &
  traffic matters \\
Sequence numbers & checking and re-arming & 2$\times$ transferred bytes &
  fixed cost dominates \\
\bottomrule
\end{tabular}
}
\end{table}

For a unicast byte account, a one-hop exchange moves approximately
$(N-1)S$ per GPU, while reduce-scatter plus all-gather moves
$2(N-1)S/N$. At $N=8$, these are $7S$ and $1.75S$; at $N=4$, they are
$3S$ and $1.5S$. At $N=2$, both move $S$, so the second phase has no byte
saving to repay its dependency cost. The two-GPU shipping policy consequently
uses one-hop paths throughout its supported range.

For two candidate protocols, the crossing of their locally fitted cost lines
identifies the switching region. \sys benchmarks the neighbouring sizes and
places the final boundary on a measured point, accounting for discrete changes
in grid shape, pipeline depth, and bandwidth saturation. At runtime, a
constant-time lookup indexed by GPU architecture, group size, and message size
returns the selected policy entry.

In the resulting policy, sequence-number readiness is used on Blackwell for
two-GPU messages up to
128\,KB. Sentinel readiness is used for the remaining shipping configurations,
including the one-hop data exchange and the boundary between reduce-scatter
and all-gather in the sharded path.

\subsection{Reduction Consistency under Asynchronous Arrival}
\label{sec:async-consistency}

Self-synchronizing transfers allow data blocks to arrive in
different orders. That freedom is harmless for data movement, but not
automatically for floating-point reduction. Floating-point addition is not
associative, so two GPUs that fold the same contributions in different arrival
orders can produce different bit patterns. A tolerance check against a
high-precision reference may accept both results, even though a subsequent
argmax, threshold, or routing decision can diverge across GPUs.

\sys separates the order in which values arrive from the order in which they
are evaluated. For replicated one-hop reductions, each contribution is first
stored in a slot identified by its source rank. Once all required slots are
ready, every GPU folds them in the same rank order, from rank 0 through
rank $N-1$. Transport remains asynchronous, but the floating-point expression
is identical across replicas.

Sharded reductions obtain the same property through ownership. Each output
block has one owner GPU, which performs the reduction once and distributes
that result to the other GPUs. The receivers do not independently recompute
the block, so their arrival order cannot create different copies. These two
cases yield the design rule used throughout \sys:
  an output block is either computed once and distributed, or every independent
replica evaluates its contributions in the same order.
This rule guarantees agreement among the participants of one collective. The
resulting value is defined by the common rank order or by the single owner's
computation, as implemented by the kernels in
Section~\ref{sec:implementation}.

\section{Implementation}
\label{sec:implementation}
\label{sec:adominates}

We implement \sys as a standalone communication backend for
SGLang~\cite{sglang2026}. The
Python runtime manages admission, symmetric storage, and policy dispatch;
architecture-specific CUDA kernels implement the data paths in
Section~\ref{sec:channels}. Hopper and Blackwell use separate kernel trees and
policy tables, while sharing the same communicator interface. The backend is
constructed once for a process group and invoked directly from the serving
runtime's collective selection chain.

\subsection{Runtime Integration and Admission}
\label{sec:runtime-integration}

The serving runtime constructs \sys for eligible process groups. During
initialization, the communicator checks that the process group runs on CUDA,
the GPU architecture has a policy table, the group lies inside a fully
connected NVLink domain, the group size is 2, 4, or 8, and the configured data
type is bf16 or fp32. It then allocates the shared storage described below and
compiles the CUDA implementation for the detected compute-capability family.

Each invocation performs a second, tensor-level admission check. The tensor
must use the communicator's device and data type, have a peer-accessible weakly
contiguous layout, satisfy 16-byte address and length alignment, remain below
the configured payload ceiling, and match an entry in the policy table for its
group size. Eligible tensors enter \sys; all other tensors continue through
the serving runtime's existing backend chain. This admission path integrates
the optimized kernels without changing communication for other layouts, data
types, or message sizes.

The runtime dispatch itself is deliberately simple. GPU compute capability
selects the Hopper or Blackwell backend, and a monotone table indexed by group
size and message size selects an algorithm identifier and fixed launch shape.
The request path performs neither fitting nor autotuning. Tables are produced
offline from the measurement procedure described in
Section~\ref{sec:protocol-selection} and evaluated in
Section~\ref{sec:results}.

\subsection{Symmetric Storage and Graph-Safe Reuse}
\label{sec:storage}

The communicator allocates one symmetric-memory slab per GPU. The allocation
provides two views of the same storage: peer pointers for unicast loads and
stores, and an NVSwitch multicast alias for \code{multimem.*}
operations~\cite{nvidia2026ptx}. The
slab is partitioned by algorithm family because one-hop publication,
reduce-scatter publication, and all-gather results require different layouts
and maximum strides. Each family receives only the capacity required by its
largest shipping policy entry.

Initialization first zeroes switch-reduction scratch regions and fills the
arrival-checked regions with the data type's sentinel. A group rendezvous and
barrier then ensure that every multicast alias is live and no peer can publish
into a region while another process is still initializing it. The communicator
performs these setup barriers once during construction. Steady-state collective
invocations begin with fully initialized aliases and storage.

Every family divides its storage into four rotating slots. Reusing a slot on
successive operations would allow a fast rank to observe a value left by an
earlier call, so each kernel derives its slot from a device-local call counter.
The counter is maintained per warp and per algorithm family. Computing the
slot on device is required for CUDA Graph execution: host launch arguments are
captured once and cannot advance a call index during graph replay. The
readiness encoding proves that a value belongs to the active protocol state;
slot rotation provides temporal separation between successive invocations.

This layout imposes two invariants. First, all ranks must issue the same
sequence of collective calls, including empty tensors, so their counters remain
aligned. Second, an algorithm family uses one fixed grid. Changing the mapping
between warps and counters would cause different threads to select different
slots. Production dispatch therefore obtains the launch shape from the same
policy entry that sized the counters and slab.

\subsection{Communication Kernels}
\label{sec:kernels}

\textbf{One-hop transfer.} For smaller messages, each GPU publishes its input
directly to the other participants through unicast peer stores or an NVSwitch
multicast store. Receive storage is indexed by source rank. A thread polls the
required source slots, resets each sentinel after capturing the value, and
performs the reduction only after all contributions are available. For more
than two GPUs, contributions are held by source rank and folded in rank order,
as described in Section~\ref{sec:async-consistency}. A warp-per-peer variant
assigns independent warps to sources, turning sequential peer waits into
concurrent polling when readiness latency dominates.

The low-latency two-GPU variant replaces sentinels with sequence-numbered
packets. Each 16-byte packet carries 8 bytes of payload and two copies of the
current sequence number. The receiver accepts the packet when the sequence
matches its call index, then reduces the payload without a sentinel reset.
Because this encoding doubles the transferred bytes, the shipping policy uses
it only through 128\,KB on Blackwell with two GPUs.

\textbf{Sharded two-hop transfer.} For larger groups and messages, the output
is divided into $N$ shards. Every GPU first publishes its input in the
reduce-scatter region. The owner of shard $r$ obtains the reduction for that
shard, either by reading through \code{multimem.ld\_reduce} or by explicitly
folding published peer values. It then distributes the completed shard through
multicast or peer stores. Receivers poll the all-gather region and place each
completed shard in the final output. Sentinel checks connect the two phases at
block granularity, so an owner can distribute one completed block while other
blocks are still waiting for input.

The kernel overlaps publication, reduction, and result collection through
bounded windows. A window deeper than the number of packs owned by a thread
degenerates into repeated dependency waits, so pipeline depth is capped by the
available work. Publication is also split between the threads reducing the
local shard and those publishing remote shards, avoiding duplicated ownership.
These launch and pipeline parameters are encoded in the offline policy table
rather than selected dynamically inside the kernel.

\textbf{All-gather.} The result-distribution half of the two-hop kernel already
implements an all-gather: each source publishes one shard and receivers wait
for the remote shards through the same readiness layer. The standalone
all-gather reuses the symmetric slab, device-side slot rotation, and sentinel
polling, omitting only the reduction step. Its performance is reported in
Section~\ref{sec:allgather}.

\subsection{Launch Dependencies and Progress}
\label{sec:progress}

All shipping kernels use Programmatic Dependent Launch
(PDL)~\cite{nvidia2026cuda} to connect to
neighbouring GPU work under CUDA Graph execution. Slot selection, offsets, and
shard indices do not depend on the input values, so \sys computes them before
the PDL wait and overlaps this front-end work with completion of the preceding
kernel. Payload reads occur only after the dependency is satisfied, and the
kernel signals dependent work after its output is published.

Production readiness loops persist until the required peer data arrives. This
requires publishing and polling blocks to remain concurrently resident, so
the adapter admits launch grids within the measured co-residency bound. For a
non-finite switch-reduction result, the kernel changes from aggregate polling
to explicit source verification after a bounded interval. It completes as soon
as every source is present, including inputs containing a legitimate NaN or
infinity.

\subsection{Steady-State Synchronization Cost}
\label{sec:barriercount}

The two-hop path provides a source-level comparison with the strongest
baselines because all three implementations use the same reduce-scatter plus
all-gather structure. Their steady-state synchronization differs as follows.
The two baseline bookends add separate cross-GPU coordination before and after
data movement. \sys establishes both dependencies through block-level
readiness evidence, removing the two phases from the steady-state path. This
reduction in fixed protocol work produces the lower latency intercept measured
in Section~\ref{sec:results}. Table~\ref{tab:barrier-count} summarizes the
resulting steady-state barrier counts.

\begin{table}[t]
\centering
\caption{\textbf{The two-hop \sys kernel removes both cross-GPU bookend
barriers.} Barrier counts cover group-wide and semaphore-style inter-GPU
coordination in the steady-state collective.}
\label{tab:barrier-count}
\fitwidth{%
\begin{tabular}{@{}p{0.26\linewidth} p{0.23\linewidth} p{0.23\linewidth} c@{}}
\toprule
\textbf{Implementation} & \textbf{Entry condition} &
\textbf{Exit condition} & \textbf{Barriers} \\
\midrule
NCCL symmetric window & LSA multimem barrier & release synchronization &
  $2$ \\
SGLang custom AR v2 & \code{barrier.arrive\_relaxed(0)} &
  \code{barrier.arrive\_rel\_acq(1)} & $2$ \\
\sys & input/output readiness evidence & block consumption and re-arming &
  $0$ \\
\bottomrule
\end{tabular}}
\end{table}

\section{Evaluation}
\label{sec:results}

We evaluate \sys at two levels. Collective microbenchmarks measure whether
self-synchronizing transfers reduce the latency of all-reduce and all-gather
across message sizes and GPU counts. We then integrate \sys into an LLM
serving stack and measure whether these savings reduce time per output token
(TPOT) during decoding.

\subsection{Experimental Setup}
\label{sec:eval-setup}

Our collective experiments use one eight-GPU node and form communicators of
two, four, and eight GPUs. We evaluate NVIDIA H100, H200, and B300 systems,
which cover two NVLink generations. The evaluation uses two complementary
views of all-reduce performance. A detailed eight-GPU B300 sweep shows how
absolute latency changes with message size, while the full cross-system sweep
shows how the median benefit changes with hardware and communicator width.
The collective experiments cover message sizes from 1\,KB to 16\,MB. Unless
stated otherwise, latency is measured inside CUDA Graph execution, where host
launch overhead does not obscure the device-side synchronization cost.

For all-reduce, the primary comparison is SGLang custom AR v2, the strongest
decode-oriented baseline over the target operating range. Both the detailed
B300 comparison and the cross-system medians use this baseline. The all-gather
experiment compares against the per-size fastest of three available baselines.
All reported improvements use lower latency as positive.

The end-to-end experiment runs DeepSeek-V4-Flash-0731 with tensor-parallel
decoding on all eight B300 GPUs. We compare the same serving workload using
\sys and custom AR v2. TPOT measures the average latency
between consecutively generated output tokens. We report concurrency 16 and
above, which covers the loaded serving regime used by this experiment.

\subsection{All-Reduce Performance}
\label{sec:threeworlds}

\begin{figure}[t]
\centering
\begin{tikzpicture}
\begin{groupplot}[
  group style={group size=2 by 1, horizontal sep=28pt},
  figbase,
  width=0.47\linewidth,
  height=4.2cm,
  xmode=log,
  log basis x=2,
  xmin=0.7, xmax=23000,
  xtick={1,8,64,512,4096,16384},
  xticklabels={1\,KB,8\,KB,64\,KB,512\,KB,4\,MB,16\,MB},
  xlabel={message size},
  ymajorgrids,
  grid style={line width=0.2pt, draw=gray!25},
  axis lines=left,
  every axis plot/.append style={line width=1.0pt, mark size=1.6pt},
]
\nextgroupplot[
  title={(a) Absolute latency},
  ylabel={latency ($\mu$s)},
  ymode=log,
  log basis y=10,
  ymin=2.8, ymax=65,
  ytick={3,5,10,20,50},
  yticklabels={3,5,10,20,50},
  legend style={at={(1.08,-0.31)}, anchor=north, legend columns=2,
                font=\scriptsize, draw=none, fill=none, column sep=8pt},
]
\addplot[black!55, mark=square*] coordinates {
  (1,3.49) (2,3.67) (4,3.72) (8,3.80) (16,3.86) (32,3.96)
  (64,4.18) (128,4.92) (256,6.42) (512,9.45) (1024,12.36)
  (2048,15.86) (4096,20.70) (8192,33.10) (16384,53.43)};
\addplot[cB300, very thick, mark=*] coordinates {
  (1,3.33) (2,3.38) (4,3.74) (8,3.84) (16,3.89) (32,3.99)
  (64,4.16) (128,4.74) (256,6.13) (512,6.58) (1024,7.08)
  (2048,9.10) (4096,14.78) (8192,23.17) (16384,42.48)};
\legend{Custom AR v2,\sys}

\nextgroupplot[
  title={(b) Latency reduction},
  ylabel={reduction (\%)},
  ymin=-10, ymax=48,
]
\draw[gray!70, dashed, line width=0.5pt]
  (axis cs:0.7,0) -- (axis cs:23000,0);
\addplot[cB300, very thick, mark=*] coordinates {
  (1,4.6) (2,7.9) (4,-0.5) (8,-1.1) (16,-0.8) (32,-0.8)
  (64,0.5) (128,3.7) (256,4.5) (512,30.4) (1024,42.7)
  (2048,42.6) (4096,28.6) (8192,30.0) (16384,20.5)};
\end{groupplot}
\end{tikzpicture}
\caption{All-reduce latency on eight B300 GPUs. Panel~(a) compares the
absolute latency of custom AR v2 and \sys; panel~(b) shows the corresponding
reduction. Points sample 15 message sizes from the 188-size sweep.}
\label{fig:b300-ar}
\end{figure}
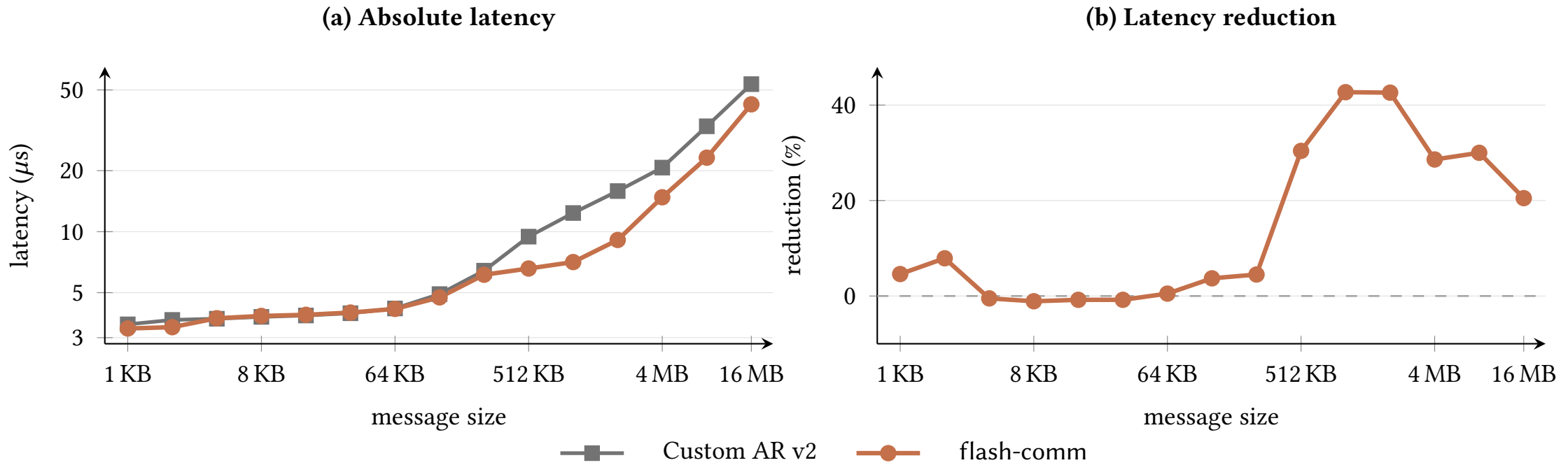

We first examine the detailed eight-GPU B300 sweep. Across all 188 message
sizes, \sys reduces median latency by 28.8\% and reaches a maximum reduction
of 43.7\%. Table~\ref{tab:ar-latency} reports representative absolute
latencies, while Figure~\ref{fig:b300-ar} shows the size-dependent trend.

\begin{table}[t]
\centering
\caption{Representative all-reduce latency on eight B300 GPUs. The reduction
is relative to custom AR v2; latency is in microseconds.}
\label{tab:ar-latency}
\fitwidth{%
\begin{tabular}{@{}r r r r@{}}
\toprule
\textbf{Message size} & \textbf{Custom AR v2} & \textbf{\sys} &
  \textbf{Reduction} \\
\midrule
1\,KB   & 3.49 & 3.33 & 4.6\% \\
128\,KB & 4.92 & 4.74 & 3.7\% \\
512\,KB & 9.45 & 6.58 & 30.4\% \\
1\,MB   & 12.36 & 7.08 & \textbf{42.7\%} \\
4\,MB   & 20.70 & 14.78 & 28.6\% \\
16\,MB  & 53.43 & 42.48 & 20.5\% \\
\bottomrule
\end{tabular}}
\end{table}

The largest gains occur once the selector moves to the
sharded two-hop path, which reduces peer-scaled transfer volume while avoiding
the two bookend barriers described in Section~\ref{sec:barriercount}. The
advantage reaches 43.7\% over the complete sweep. At small messages, all
implementations use low-overhead one-hop protocols and fixed kernel work
dominates. Consequently, \sys is within approximately 1\% of custom AR v2 at
the weakest eight-GPU points, including a small 4--16\,KB loss region.

The absolute latency comparison gives the same picture. On eight B300 GPUs,
the implementations remain close through the small-message range. At 512\,KB,
custom AR v2 takes 9.45\,$\mu$s while \sys takes 6.58\,$\mu$s. At 16\,MB,
the corresponding latencies are 53.43 and 42.48\,$\mu$s. Thus the median gain
does not come from a uniform shift across all sizes. It comes primarily from
the region where removing separate synchronization and reducing transferred
bytes both matter.

\subsection{All-Gather Performance}
\label{sec:allgather}

\sys implements both all-gather and all-reduce using the same
self-synchronizing transfer design. The all-gather implementation
uses symmetric storage, slot rotation, and block-level readiness checks to
exchange each rank's shard among the participating GPUs.
Table~\ref{tab:ag-summary} reports representative absolute latencies on B300
against the fastest baseline at each tested size. \sys improves all 45
measured combinations of message size and GPU count. Across eight GPUs, it
reduces latency by a median of 23.8\% and by as much as 58.9\%.
Because all-gather exchanges shards without a reduction stage, its improvement
comes directly from lowering fixed coordination cost and allowing each
receiver to validate the blocks it needs. These results show that the same
self-synchronizing transfer design benefits both all-gather and all-reduce.

\begin{table}[t]
\centering
\caption{Representative all-gather latency on eight B300 GPUs. The baseline
is the fastest available implementation at each message size; latency is in
microseconds.}
\label{tab:ag-summary}
\fitwidth{%
\begin{tabular}{@{}r r r r@{}}
\toprule
\textbf{Message size} & \textbf{Fastest baseline} & \textbf{\sys} &
  \textbf{Reduction} \\
\midrule
1\,KB   & 3.36 & 2.56 & 23.8\% \\
128\,KB & 3.80 & 2.86 & 24.7\% \\
512\,KB & 8.44 & 3.47 & \textbf{58.9\%} \\
1\,MB   & 9.34 & 4.24 & 54.6\% \\
4\,MB   & 13.96 & 8.89 & 36.3\% \\
16\,MB  & 31.70 & 28.91 & 8.8\% \\
\bottomrule
\end{tabular}}
\end{table}

\subsection{Cross-Generation Calibration}
\label{sec:acrossgen}

The relative costs of unicast, multicast, and switch-assisted reduction vary
across NVLink generations, which can shift the protocol switching boundaries.
We therefore examine how the cost-guided policy adapts across hardware
platforms. Appendix~\ref{app:constants} lists the machine constants used for
calibration.

Table~\ref{tab:ar-summary} reports the median reduction
over all 188 message sizes for each configuration. All three systems obtain
their largest benefit at eight GPUs, where the reduction reaches
22.3--28.8\%. The magnitude nevertheless differs across machines, including
between H100 and H200 within the same NVLink generation.

\begin{table}[t]
\centering
\caption{Median all-reduce latency reduction relative to custom AR v2 across
188 message sizes. The benefit grows with the number of participating GPUs.}
\label{tab:ar-summary}
\fitwidth{%
\begin{tabular}{@{}l r r r@{}}
\toprule
\textbf{System} & \textbf{2 GPUs} & \textbf{4 GPUs} & \textbf{8 GPUs} \\
\midrule
H100 (NVLink 4) & 8.6\% & 11.8\% & 24.3\% \\
H200 (NVLink 4) & 4.9\% & 9.5\% & 22.3\% \\
B300 (NVLink 5) & 4.2\% & 9.8\% & \textbf{28.8\%} \\
\bottomrule
\end{tabular}}
\end{table}

The same protocol boundary therefore cannot be copied unchanged across GPU
generations or neighbouring systems. On H100 and H200, multicast throughput
is approximately one third of unicast throughput, while on B300 the two are
comparable. These different slopes and fixed costs change which data path is
efficient and where two fitted protocol costs cross. Accordingly, \sys uses
architecture-specific policy tables calibrated from the cost model in
Section~\ref{sec:protocol-selection}. The selector uses the analytical crossing
only to locate the search region and places the runtime boundary on a
neighbouring measured point.

\subsection{End-to-End Decoding}
\label{sec:e2e}

Table~\ref{tab:e2e-tpot} reports TPOT from concurrency 16 to 512. \sys has the
lowest TPOT at all ten reported points against custom AR v2.
Relative to custom AR v2, it reduces TPOT by 1.7--6.8\%. The largest relative
gains occur at concurrency 32--64, where TPOT falls by 4.8--6.8\%. The benefit
remains positive at higher concurrency, including 5.2\% at concurrency 192 and
2.2--2.9\% from concurrency 256 to 512.
The relative and absolute savings emphasize different operating points. At
concurrency 32, \sys achieves the largest relative reduction, lowering TPOT
from 4.38 to 4.08\,ms. At concurrency 192, it achieves the largest absolute
reduction, lowering TPOT from 13.60 to 12.89\,ms, a saving of 0.71\,ms per
output token. The improvement is therefore not monotonic in either concurrency
or baseline TPOT, consistent with the workload exercising different collective
sizes and policy bands as concurrency changes. Because the model and serving
configuration remain fixed across backends, the consistent lead across the
full range shows that the collective-level improvements carry through to the
decoding path under multiple load regimes.

\begin{table}[t]
\centering
\caption{\textbf{Flash-comm lowers end-to-end TPOT by 1.7--6.8\% relative to
custom AR v2 across all reported concurrency levels.} TPOT is in
milliseconds; lower is better.}
\label{tab:e2e-tpot}
\fitwidth{%
\begin{tabular}{@{}r r r r r@{}}
\toprule
\textbf{Concurrency}  & \textbf{Custom AR v2} &
  \textbf{\sys} & \textbf{Reduction vs. v2} \\
\midrule
16   & 2.99 & \textbf{2.94} & 1.7\% \\
32   & 4.38 & \textbf{4.08} & \textbf{6.8\%} \\
48   & 5.45 & \textbf{5.16} & 5.3\% \\
64   & 6.49 & \textbf{6.18} & 4.8\% \\
96   & 8.32 & \textbf{8.06} & 3.1\% \\
128  & 10.15 & \textbf{9.88} & 2.7\% \\
192  & 13.60 & \textbf{12.89} & 5.2\% \\
256  & 9.92 & \textbf{9.70} & 2.2\% \\
384  & 9.94 & \textbf{9.65} & 2.9\% \\
512  & 10.01 & \textbf{9.76} & 2.5\% \\
\bottomrule
\end{tabular}}
\end{table}

\section{Discussion and Limitations}
\label{sec:discussion}
\label{sec:boundary}

\subsection{Deployment Scope}

\sys targets collective communication within a single high-bandwidth GPU
domain. The current implementation supports communicators of two, four, and
eight GPUs connected through NVLink and NVSwitch. This scope matches the
single-node tensor-parallel groups commonly used during LLM decoding, where
collectives recur at every layer and generated token. Extending the design
across nodes would introduce network transports, routing, and failure handling
with different readiness and cost structures.

The optimized kernels require Hopper or Blackwell GPUs and use CUDA features
including symmetric peer-accessible memory, \code{multimem.*} operations, and
Programmatic Dependent Launch. The runtime admits bf16 and fp32 tensors with
peer-accessible layouts and 16-byte alignment. Other devices, data types, and
layouts continue through the serving runtime's existing communication
backends. These requirements allow readiness evidence and payload data to
share the same transfer path while preserving CUDA Graph execution.

\subsection{Message Range}

The current policy tables cover messages from 16 bytes through 16\,MB. This
range includes the small and medium collectives that dominate tensor-parallel
decoding, from short control and partial-result transfers to multi-megabyte
hidden-state exchanges at higher concurrency. The evaluation focuses on
1\,KB--16\,MB, where fixed synchronization cost and transferred bytes both
materially affect latency. Messages outside the optimized policy range use the
next communication backend selected by the serving runtime.

The supported range is also where cost-guided selection is most useful. For
short messages, fixed readiness work dominates and compact sequence-number
protocols can be effective. As messages grow, their additional bytes become
more expensive, favouring sentinel readiness. At still larger sizes and wider
GPU groups, sharded two-hop communication reduces peer-scaled traffic enough
to repay its additional phase. A single protocol therefore cannot cover the
decode range efficiently.

\subsection{Recalibration across GPU Generations}

The selection method transfers across GPU generations, while its calibrated
boundaries remain hardware specific. The intercept and slope in
Equation~\ref{eq:cost-model} depend on kernel launch cost, barrier latency,
peer-access latency, and the relative throughput of unicast, multicast, and
switch-assisted reduction. These quantities change at different rates as GPU
and interconnect generations evolve. For example, multicast is substantially
slower than unicast on the evaluated Hopper systems but reaches comparable
throughput on B300. The same communication path consequently crosses its
alternatives at a different message size.

\sys handles this variation through architecture-specific policy tables. For
each platform, offline measurements fit the fixed and per-byte costs of the
candidate protocols, locate their crossing regions, and benchmark neighbouring
message sizes to select measured boundaries. Runtime dispatch then reduces to
a table lookup. This separation preserves the main design principle across
hardware generations: the cost model and calibration procedure remain stable,
while the selected readiness encoding, data path, and switching points adapt
to the target system.

\section{Related Work}
\label{sec:related-work}

\subsection{Low-Latency GPU Collectives}

NCCL provides architecture-tuned collective algorithms and protocols across a
wide range of topologies and message sizes~\cite{nvidia2026nccl}. Prior analyses have exposed
how its protocol, channel, and algorithm choices determine latency and
bandwidth~\cite{hu2025demystifying}. More recent systems push small-message collectives closer to
the hardware limit. Every Microsecond Matters specializes GPU collectives for
latency-sensitive workloads and carefully accounts for kernel and transport
overheads~\cite{shen2026every}. SiFAR targets inference all-reduce and removes global
synchronization through a synchronization-free data path on H200~\cite{taneja2026sifar}. Studies
of the synchronization tax further quantify how coordination limits GPU
scale-up performance across workloads and hardware generations~\cite{devraj2026synchronization}.

\sys shares the goal of minimizing exposed collective latency, but focuses on
how data readiness is represented and consumed. Its self-synchronizing
transfers attach readiness evidence to each block, allowing the receiver to
advance at block granularity and re-arm storage through the same protocol. The
cost-guided selector then chooses among readiness encodings and communication
paths according to their fixed and per-byte costs. This combination extends
the mechanism from all-reduce to all-gather and makes protocol boundaries an
explicit, hardware-calibrated decision.

Asynchronous arrival also creates a numerical concern that is separate from
ordinary approximation error. Floating-point addition is not associative, so
replicas that reduce contributions in arrival order can diverge even when they
receive identical inputs~\cite{ieee2019floating}. \sys preserves cross-GPU agreement by assigning
a common source-rank order to replicated reductions and a single owner to each
sharded output block. Transport can therefore remain asynchronous without
making the result depend on arrival timing.

\subsection{Communication Optimization for LLM Systems}

Another line of work reduces exposed communication by overlapping it with
computation. TokenWeave schedules distributed inference to increase
compute-communication overlap~\cite{gond2025tokenweave}. Comet and tile-level signaling systems
coordinate communication and computation at fine granularity for
mixture-of-experts workloads~\cite{zhang2025comet,cui2026tile}, while Domino applies tensor slicing and
overlap to distributed transformer execution~\cite{wang2024domino}. X-Stage models the
backpressure between communication issue and remote visibility to improve
fusion decisions~\cite{xian2026xstage}. These systems change when communication executes or
how much of it is visible on the critical path.

\sys instead reduces the latency of the collective operation itself. This is
particularly important during decoding, where a dependent layer often waits
for the collective result and the short per-token computation window limits
the amount of communication that can be hidden. The two approaches are
complementary: scheduling and fusion can overlap the portion permitted by the
dependency graph, while self-synchronizing transfers reduce the remaining
transport and coordination cost.

\subsection{Scaling and Hardware-Aware Selection}

Large-scale collective systems optimize routing, topology, and bandwidth for
multi-node deployments containing thousands of GPUs~\cite{si2025collective,singhania2025multinode}. Their dominant
costs differ from the single-node decode regime studied here, where repeated
small and medium transfers make fixed synchronization latency visible. \sys
therefore specializes within one NVLink domain and leaves inter-node routing
and transport to the existing communication stack.

Hardware-aware algorithm selection is already fundamental to collective
libraries. Our measurements show that readiness protocols require the same
treatment: barrier latency and the relative throughput of unicast, multicast,
and switch-assisted reduction change across GPU generations. \sys captures
these changes through the intercept and slope of its cost model, then
calibrates the switching points with neighbouring measurements. The portable
artifact is the selection method, while the resulting policy table is tied to
the target GPU and interconnect generation.

\section{Conclusion}
\label{sec:conclusion}

Collective communication remains on the critical path of distributed LLM
decoding, even on modern high-bandwidth GPU systems. Our measurements identify
separate readiness synchronization as an important source of this cost.

\sys removes these phases through self-synchronizing transfers that fuse
readiness synchronization with data movement. Each block carries evidence of
its own readiness, allowing receivers to validate, consume, and safely reuse
storage without a separate group-wide barrier. A cost-guided selector balances
the fixed work of readiness handling against additional transferred bytes and
chooses among readiness encodings, one-hop exchange, and sharded two-hop
communication. Finally, \sys decouples transport order from reduction order,
preserving cross-GPU agreement while allowing blocks to arrive asynchronously.

On eight B300 GPUs, \sys reduces all-reduce latency by a median of 28.8\% and
up to 43.7\% relative to custom AR v2. 
The same mechanism reduces all-gather latency by a median of 23.8\% and up to 58.9\%. 
In end-to-end decoding, these collective improvements reduce TPOT by 1.7--6.8\%.
The principle also applies to all-to-all,
which has similar peer-readiness dependencies without a reduction step. We
plan to adapt \sys to all-to-all and evaluate its impact on expert-parallel
workloads.
More importantly, our extensive experiments demonstrate that the benefits of this design 
generalize across a wide range of message sizes and GPU generations.
Together, these results provide a practical path for
keeping decode communication efficient as GPU systems continue to evolve.

\begingroup
\small
\bibliographystyle{unsrt}
\bibliography{references}

@article{shen2026every,
  author  = {Siyuan Shen and Anton Korzh and John Bachan and Tiancheng Chen and
             Arnav Goel and Ludwig Schneider and Pouya Kousha and Zhenhao He and
             Sylvain Jeaugey and Kamil Iskra and Nishank Chandawala and
             Jeff R. Hammond and Torsten Hoefler},
  title   = {Every Microsecond Matters: Achieving Near-Speed-of-Light Latency in
             {GPU} Collectives},
  journal = {arXiv preprint arXiv:2607.16100},
  year    = {2026},
}

@article{taneja2026sifar,
  author  = {Hritvik Taneja and Anish Saxena and Abhishek Revinipati and
             Jae Hyung Ju and Neal C. Crago and Moinuddin Qureshi},
  title   = {{SiFAR}: Synchronization-Free All-Reduce for Low-Latency {LLM}
             Inference},
  journal = {arXiv preprint arXiv:2607.08973},
  year    = {2026},
}

@article{hu2025demystifying,
  author  = {Zhiyi Hu and Siyuan Shen and Tommaso Bonato and Sylvain Jeaugey and
             Cedell Alexander and Eric Spada and James Dinan and Jeff Hammond and
             Torsten Hoefler},
  title   = {Demystifying {NCCL}: An In-depth Analysis of {GPU} Communication
             Protocols and Algorithms},
  journal = {arXiv preprint arXiv:2507.04786},
  year    = {2025},
}

@article{devraj2026synchronization,
  author  = {Arjun Devraj and Lindsey Bowen and Rachee Singh},
  title   = {Understanding the Synchronization Tax in {GPU} Scale-Up Domains},
  journal = {arXiv preprint arXiv:2608.22503},
  year    = {2026},
}

@article{si2025collective,
  author  = {Min Si and Pavan Balaji and Yongzhou Chen and Ching-Hsiang Chu and
             Adi Gangidi and Saif Hasan and Subodh Iyengar and Dan Johnson and
             Bingzhe Liu and Regina Ren and Deep Shah and Ashmitha Jeevaraj Shetty
             and Greg Steinbrecher and Yulun Wang and Bruce Wu and Xinfeng Xie and
             Jingyi Yang and Mingran Yang and Kenny Yu and Minlan Yu and Cen Zhao
             and others},
  title   = {Collective Communication for 100k+ {GPUs}},
  journal = {arXiv preprint arXiv:2510.20171},
  year    = {2025},
}

@article{singhania2025multinode,
  author  = {Prajwal Singhania and Siddharth Singh and Lannie Dalton Hough and
             Akarsh Srivastava and Harshitha Menon and Charles Fredrick Jekel and
             Abhinav Bhatele},
  title   = {Understanding and Improving Communication Performance in
             Multi-node {LLM} Inference},
  journal = {arXiv preprint arXiv:2511.09557},
  year    = {2025},
}

@article{gond2025tokenweave,
  author  = {Raja Gond and Nipun Kwatra and Ramachandran Ramjee},
  title   = {{TokenWeave}: Efficient Compute-Communication Overlap for
             Distributed {LLM} Inference},
  journal = {arXiv preprint arXiv:2505.11329},
  year    = {2025},
}

@article{zhang2025comet,
  author  = {Shulai Zhang and Ningxin Zheng and Haibin Lin and Ziheng Jiang and
             Wenlei Bao and Chengquan Jiang and Qi Hou and Weihao Cui and
             Size Zheng and Li-Wen Chang and Quan Chen and Xin Liu},
  title   = {{Comet}: Fine-grained Computation-Communication Overlapping for
             Mixture-of-Experts},
  journal = {arXiv preprint arXiv:2502.19811},
  year    = {2025},
}

@article{wang2024domino,
  author  = {Guanhua Wang and Chengming Zhang and Zheyu Shen and Ang Li and
             Olatunji Ruwase},
  title   = {{Domino}: Eliminating Communication in {LLM} Training via Generic
             Tensor Slicing and Overlapping},
  journal = {arXiv preprint arXiv:2409.15241},
  year    = {2024},
}

@article{cui2026tile,
  author  = {Minyu Cui and Anna Wingkvist and Morgan Ericsson},
  title   = {Fine-grained Computation-Communication Overlap via Tile-level
             Signaling and Scheduling for Mixture-of-Experts},
  journal = {arXiv preprint arXiv:2607.19539},
  year    = {2026},
}

@article{xian2026xstage,
  author  = {Jianwen Xian and Zhiyuan Xu and Yuchen Li and Ziliang Lai and
             Kang He and Zhen Huang and Aichen Feng and Jinyan Chen and
             Yilin Zhang and Qinqin Chen and Julien Lai and Chengru Song},
  title   = {{X-Stage}: Modeling Post-Issue Backpressure in {GPU}
             Communication--Computation Fusion},
  journal = {arXiv preprint arXiv:2607.23264},
  year    = {2026},
}

@manual{nvidia2026ptx,
  author       = {{NVIDIA Corporation}},
  title        = {Parallel Thread Execution {ISA}},
  organization = {NVIDIA Corporation},
  edition      = {9.0},
  year         = {2026},
  note         = {Documentation for \texttt{multimem} instructions},
}

@misc{nvidia2026nccl,
  author       = {{NVIDIA Corporation}},
  title        = {{NVIDIA Collective Communications Library (NCCL)}},
  year         = {2026},
  howpublished = {\url{https://github.com/NVIDIA/nccl}},
}

@manual{nvidia2026cuda,
  author       = {{NVIDIA Corporation}},
  title        = {{CUDA C++ Programming Guide}},
  organization = {NVIDIA Corporation},
  edition      = {13.0},
  year         = {2026},
  note         = {Programmatic Dependent Launch and synchronization},
}

@misc{ieee2019floating,
  author       = {{IEEE Computer Society}},
  title        = {{IEEE Standard for Floating-Point Arithmetic}},
  organization = {IEEE},
  number       = {IEEE Std 754-2019},
  year         = {2019},
}

@misc{sglang2026,
  author       = {{LMSYS Org}},
  title        = {{SGLang}: A Fast Serving Framework for Large Language Models
                  and Vision Language Models},
  year         = {2026},
  howpublished = {\url{https://github.com/sgl-project/sglang}},
}
\endgroup

\appendix
\appendixformat
\section{Platform Calibration Measurements}
\label{app:constants}

Table~\ref{tab:platform-calibration} reports the measurements used to
calibrate Equation~\ref{eq:cost-model}. Latency quantities contribute
primarily to the fixed term $a_p$, while data-path throughput determines the
per-byte term $b_p$. All systems use the same probe definitions, and latency
probes execute inside CUDA Graphs. Fixed-shape throughput rows use an identical
launch shape across systems; ceiling rows sweep launch shapes and report the
best observed result.

\begin{table}[t]
\centering
\caption{\textbf{Synchronization and data-path costs change differently across
GPU generations.} Latency probes run inside CUDA Graphs; throughput is reported
as effective GB/s.}
\label{tab:platform-calibration}
{\small
\fitwidth{%
\begin{tabular}{@{}l r r r l@{}}
\toprule
\textbf{Quantity} & \textbf{H100} & \textbf{H200} & \textbf{B300} &
  \textbf{Unit} \\
\midrule
\multicolumn{5}{@{}l}{\textit{Fixed and latency costs}} \\
Empty-kernel floor & 0.790 & 0.507 & 0.525 & $\mu$s \\
Net cross-GPU barrier & 2.66 & 1.87 & 4.33 & $\mu$s \\
1\,KB all-reduce, 8 GPUs & 2.54 & 2.46 & 3.33 & $\mu$s \\
Barrier / 1\,KB all-reduce & 1.05 & 0.76 & 1.30 & ratio \\
L2 cross-side latency gap & 44.6 & 44.0 & 93.6 & ns \\
\addlinespace[2pt]
\multicolumn{5}{@{}l}{\textit{Data-path throughput}} \\
Unicast store, fixed shape @256\,MB & 287.3 & 322.6 & 261.1 & GB/s \\
Multicast store, fixed shape @256\,MB & 98.4 & 98.3 & 260.8 & GB/s \\
Unicast store, shape-swept ceiling & 359.9 & 359.3 & 652.6 & GB/s \\
Multicast team egress, 8 GPUs & 363.2 & 363.2 & 709.3 & GB/s \\
All-to-all egress, 8 GPUs & 334.4 & 334.8 & 636.2 & GB/s \\
Peer \code{red.add} & 324.4 & 324.8 & 558.8 & GB/s \\
Switch \code{multimem.red} & 98.4 & 98.1 & 237.7 & GB/s \\
Local reduction & 840.8 & 1128.7 & 2337.9 & GB/s \\
Local HBM copy & 2712.1 & 3659.3 & 6634.8 & GB/s \\
\bottomrule
\end{tabular}}
}
\end{table}

Two cross-generation changes directly affect protocol selection. First, a
standalone barrier is 0.76$\times$ the latency of a 1\,KB all-reduce on H200,
but rises to 1.30$\times$ on B300. Second, under the fixed-shape probe,
multicast store throughput is 0.34$\times$ and 0.30$\times$ unicast on H100
and H200, respectively, but reaches 1.00$\times$ on B300. Blackwell therefore
changes both the value of removing a synchronization phase and the relative
cost of multicast publication.

The systems also differ in the resources surrounding the fabric. H100 and
H200 provide 132 SMs and 50.0/60.0\,MiB of L2, while B300 provides 148 SMs and
126.5\,MiB of L2. The eight-GPU reduce-scatter/all-gather overlap probe reaches
68\% on H100 and H200 and 52\% on B300; the peer-write probe observes no torn
128-byte writes on any system. \sys combines these measured fixed and per-byte
costs to locate each protocol crossing, benchmarks the neighbouring message
sizes, and installs the resulting boundary in the architecture-specific policy
table.

\end{document}